\documentclass[journal]{IEEEtran}
\pdfoutput=1
\usepackage[latin1]{inputenc}
\usepackage{graphicx}% Include figure files
\usepackage{dcolumn}% Align table columns on decimal point
\usepackage{bm}% bold math
\usepackage{units}

\usepackage{color}
\usepackage{hyperref}
\begin{document}
\title{Nanofiber-based optical trapping \\ of cold neutral atoms}% Force line breaks with \\

\author{E.~Vetsch,
				S.~T.~Dawkins,
				R.~Mitsch,
				D.~Reitz,
				P.~Schneeweiss,
				and ~A.~Rauschenbeutel
        }%

\maketitle

\begin{abstract}
We present experimental techniques and results related to the optimization and characterization of our nanofiber-based atom trap [Vetsch {\it et al.}, Phys. Rev. Lett. {\bf 104}, 203603 (2010)]. The atoms are confined in an optical lattice which is created using a two-color evanescent field surrounding the optical nanofiber. For this purpose, the polarization state of the trapping light fields has to be properly adjusted. We demonstrate that this can be accomplished by analyzing the light scattered by the nanofiber. Furthermore, we show that loading the nanofiber trap from a magneto-optical trap leads to sub-Doppler temperatures of the trapped atomic ensemble and yields a sub-Poissonian distribution of the number of trapped atoms per trapping site.

%\tableofcontents

\end{abstract}
%\IEEEpeerreviewmaketitle

%#############################################################################################
%#############################################################################################
%#############################################################################################
\section{Introduction}
%\setstcolor{red}
\footnotetext[2012]{\copyright $ $ IEEE, see \url{http://www.ieee.org} for copyright policy \\IEEE J. Quantum Electron. 18, 1763 (2012)\\ DOI: 10.1109$/$JSTQE.2012.2196025}
We have recently demonstrated a new experimental platform for the simultaneous trapping and optical interfacing of laser-cooled cesium atoms \cite{Vetsch10, Dawkins11}. The scheme uses a two-color evanescent field surrounding an optical nanofiber to localize the atoms in a one-dimensional optical lattice about 200~nm above the nanofiber surface \cite{LeKien04}, see Fig.~\ref{fig:setup}. At the same time, the atoms can be efficiently interrogated with light which is sent through the nanofiber. Remarkably, an ensemble of 2000 trapped atoms almost entirely absorbs a resonant probe field, yielding an optical depth of up to 32, equivalent to an absorbance per atom of 1.6~\% \cite{Rauschenbeutel11}. Moreover, when dispersively interfacing the atoms, we observe a phase shift per atom of $\sim 1$~mrad at a detuning of six times the natural linewidth \cite{Dawkins11}. Our technique opens the route towards the direct integration of laser-cooled atomic ensembles within fiber networks, an important prerequisite for large-scale quantum communication. Moreover, our nanofiber trap is ideally suited to the realization of hybrid quantum systems that combine atoms with solid state quantum devices \cite{Hafezi11}.

On the route towards the realization of our nanofiber-based atom trap, two important experimental questions had to be addressed: First, the overall optical potential critically depends on the polarization state of the nanofiber-guided trapping fields \cite{LeKien04}. However, the nanofiber is realized as the waist of a tapered optical fiber which is non-polarization maintaining for technical reasons. In order to compensate the birefringence of the optical fiber sections which are guiding the trapping light into the nanofiber waist, we thus had to develop a method for measuring the polarization state of the light field inside the nanofiber itself. It turns out that this is possible by analyzing the light scattered out of the nanofiber for different input polarizations. The second experimental challenge was to load laser-cooled atoms into the potential minima. In general, the most straightforward method for loading atoms into an optical dipole trap is to directly transfer them from a magneto-optical trap (MOT), possibly while further cooling the atoms with an optical molasses~\cite{Kuppens00}. It was, however, far from obvious that this procedure would work for our nanofiber trap since it implied superposing the MOT with the nanofiber and required the MOT to efficiently cool the atoms at a distance of only $\approx 200$~nm from the nanofiber surface. In spite of these unknowns, the technique turns out to be successful. In particular, it is possible to prepare nanofiber-trapped ensembles with sub-Doppler temperatures. Moreover, the loading process leads to a sub-Poissonian distribution of the number of atoms per trapping site due to the so-called collisional blockade effect~\cite{Schlosser02}. Here, we present the experimental techniques and results related to the characterization of the polarization state of the nanofiber-guided trapping light fields, of the temperature of the nanofiber-trapped atoms, and of the number of atoms per trapping site.

\section{Experimental Set-up}
\begin{figure}
	\includegraphics{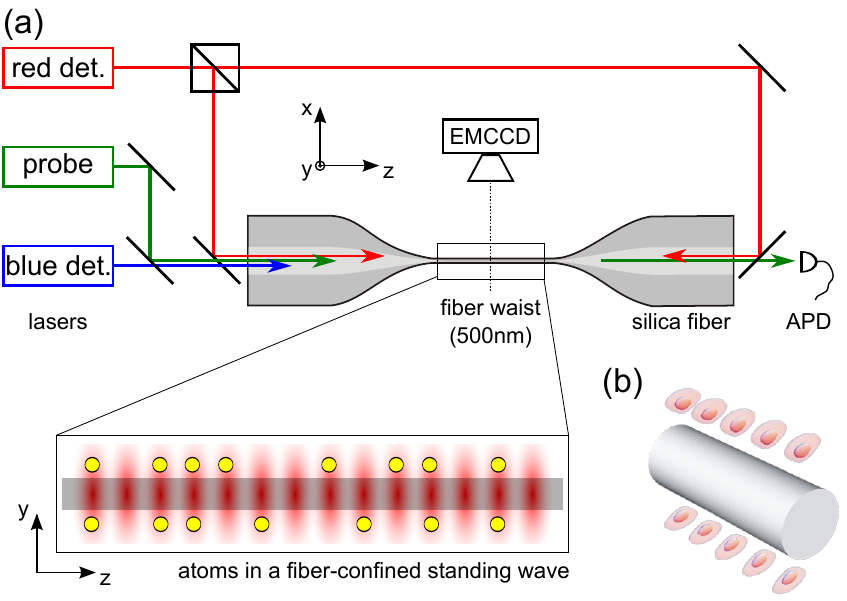}
	\caption{Illustration of the experimental system: (a) shows a sketch of the silica nanofiber, the trapping laser fields, the probe laser field, an electron multiplying charge coupled device (EMCCD) camera used for imaging the atomic fluorescence, and an avalanche photodiode (APD) employed for measuring the probe light transmitted through the fiber. The zoom shows the atoms in the nanofiber trap, including the evanescent standing wave pattern formed by the red-detuned trapping field. In (b), equipotential surfaces $\unit[40]{\mu K}$ and $\unit[125]{\mu K}$ above the trapping minimum are shown. This configuration results in two linear arrays of local potential minima (trapping sites).}
	\label{fig:setup}
\end{figure}
All measurements presented in this article are carried out with the experimental set-up which is depicted in Fig.~\ref{fig:setup}~(a) and which was also employed for the work in Ref.~\cite{Vetsch10}. A silica fiber, having a $\unit[5]{mm}$ long tapered region of $\unit[500]{nm}$ diameter, is placed in an ultra-high vacuum chamber at a typical pressure of $\unit[8 \times 10^{-10}]{mbar}$. To realize the two-color nanofiber dipole trap for Cesium atoms, two light fields are launched into the fiber. One has a free-space wavelength of $\lambda_0^{\rm blue}=\unit[780]{nm}$ and is thus blue-detuned with respect to the Cesium D-line transitions, $\lambda_0^{\rm D1}=\unit[894]{nm}$ and $\lambda_0^{\rm D2}=\unit[852]{nm}$. The second ($\lambda_0^{\rm red}=\unit[1064]{nm}$) is red-detuned and is sent into the thin fiber from both sides. In the 500-nm diameter region of the fiber, all light fields used in the presented experiment are guided in the fundamental ${\rm HE}_{11}$ mode~\cite{LeKien04b}. Both trapping light fields are chosen to have quasi-linear polarization where the polarization planes are oriented perpendicular with respect to each other in order to maximize the azimuthal confinement in the trap~\cite{Vetsch10}. The fiber-guided 780-nm mode has a power of $\unit[25]{mW}$, the respective 1064-nm standing wave field has a power of $\unit[2 \times 2.2]{mW}$. Using these parameters, two arrays of potential minima form along the nanofiber. The situation is illustrated in Fig.~\ref{fig:setup}~(b), showing equipotential surfaces at $\unit[40]{\mu K}$ and $\unit[125]{\mu K}$ above the trapping minimum. Each individual trapping site is characterized by the parameters summarized in Table~\ref{tab:Trap}. The spacing between two neighboring lattice sites is given by $\lambda_0^{\rm red}/2n_{\rm eff} \approx \unit[500]{nm}$, where $n_{\rm eff}$ is the effective refractive index \cite{Yariv07}. The closest geometrical distance to the nearest opposing trapping site of the second array of potential minima is about $\unit[1]{\mu m}$.
\begin{table}
	\begin{center}
		\begin{tabular}{|l|c|c|}
			\hline
			physical quantity & symbol & value \\
			\hline \hline
			radial trap frequency & $\nu_r$ & $\unit[200]{kHz}$ \\ \hline
			axial trap frequency & $\nu_z$ & $\unit[315]{kHz}$ \\ \hline
			azimuthal trap frequency & $\nu_\varphi$ & $\unit[140]{kHz}$ \\ \hline
			trap depth & $U_0$ & $k_{\rm B} \cdot \unit[400]{\mu K}$ \\ \hline
			trap-surface separation & $d$ & $\unit[230]{nm}$ \\ \hline
		\end{tabular}
		\caption{Overview of the parameters characterizing the trapping sites of the nanofiber trap. The constant $k_{\rm B}$ denotes Boltzmann's constant.}
		\label{tab:Trap}
	\end{center}
\end{table}

\section{Polarization Control}
Polarization control of the fiber-guided modes is an indispensable requirement in the context of nanofiber-based optical trapping. The nanofiber is produced from a non-polarization maintaining base fiber for technical reasons. Thus, the birefringence of the optical fiber sections which are guiding the trapping light into the nanofiber waist is a priori unknown. However, in order to obtain azimuthal confinement of the atoms in the nanofiber dipole trap, a defined quasi-linear polarization of the trapping laser fields is required~\cite{Vetsch10}. Therefore, a method for measuring the polarization state of the light field inside the nanofiber itself has to be implemented.

Here, we realize this characterization by analyzing the light scattered out of the nanofiber by surface imperfections and by inhomogeneities in the bulk silica. We assume that the scatterers are point-like and that they conserve the local polarization and phase of the nanofiber-guided modes. We therefore further assume that the scattered intensity in a given azimuthal direction is given by the coherent sum of the amplitudes of these local dipole emitters which are randomly distributed over the nanofiber surface and bulk.

\begin{figure}
	\includegraphics{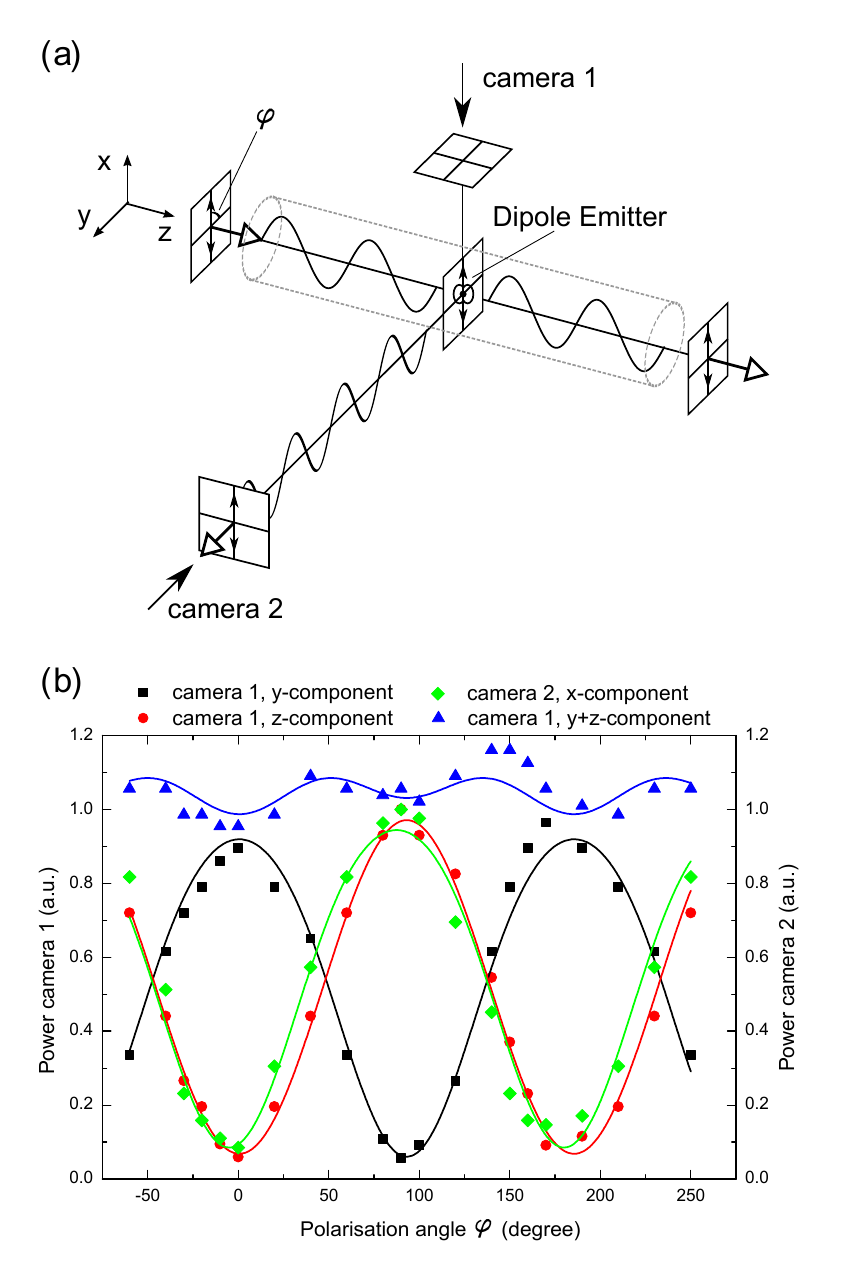}
	\caption{Characterization and optimization of the polarization state of the nanofiber-guided light. In (a), we consider one point-like scatterer which is assumed to conserve the local polarization and phase of the mode guided by the nanofiber (sketched as a dashed cylinder). If the guided light is locally linearly polarized along $x$, the scattered light will thus also be polarized along $x$ and its intensity will obey a dipole emission pattern. Two cameras allow one to image the light scattered into the $x$- and $y$-direction, respectively. Both cameras can be equipped with a polarization filter. The orientation of the polarization plane of the fiber-guided mode is given by the angle $\varphi$ as indicated. In (b), the optical powers detected by both cameras are shown as a function of $\varphi$ for various settings of the polarization filters. The solid lines are fits taking experimental imperfections into account (see text for details).}
	\label{fig:Pol}
\end{figure}

The set-up and procedure for measuring and optimizing the polarization state of the nanofiber-guided light are illustrated in Fig.~\ref{fig:Pol}~(a). The nanofiber axis defines the $z$-axis of the coordinate system. Two CCD cameras observe the scattered light along the $x$-axis (camera 1) and $y$-axis (camera 2), respectively. The apertures of the imaging optics are small, thereby ensuring a high angular resolution of about $5^\circ$.

If the mode was exactly linearly polarized along $x$, the azimuthal dependence of the scattered intensity would, under the above assumptions, be given by a dipole emission pattern, yielding $I(\varphi) \propto \sin^2(\varphi-\varphi_0)$ with $\varphi$ as defined in Fig.~\ref{fig:Pol}~(a) and $\varphi_0=90^\circ$ ($\varphi_0=0^\circ$) for camera 1 (2). However, the ${\rm HE}_{11}$ modes have a non-trivial, spatially varying polarization. As a result, due to the two-fold symmetry of the minor component of the transverse polarization~\cite{LeKien04b}, a higher order term $\propto \sin^2[2(\varphi-\varphi_0)]$ will contribute to the observed intensity. Moreover, they are not transversally polarized but exhibit a significant $z$-component. In order to nonetheless be able to adjust the polarization state of the fiber-guided modes, both CCD cameras are equipped with a polarization filter. By these means, the contribution of the $z$-component can be suppressed.

In Fig.~\ref{fig:Pol}~(b), the integrated intensity of the CCD signals of cameras 1 and 2 are plotted for various angles $\varphi$. The latter is adjusted by a half wave retardation plate before coupling the light field into the fiber. As expected, with the $z$-polarization component filtered out, the power on camera 1 (plotted in black) varies $\approx 90^\circ$ out of phase with respect to the power on camera 2 (plotted in green). We optimize the contrast of the signals using a Berek polarization compensator before coupling the light field into the fiber. In the optimum case, this yields a contrast of $C=88$~\%, most probably limited by non-polarization maintaining scattering. We checked that the light exiting the tapered optical fiber has a pure polarization state which is stable over many hours. Therefore, we infer that by maximizing $C$, we realize the best possible approximation of a quasi-linearly polarized HE$_{11}$ mode on the nanofiber waist.

While the above discussion only considered the optimization of the polarization of the red-detuned trapping light field at a wavelength of 1064~nm, it turns out that the single Berek compensator suffices to also compensate the birefringence of the fiber for the other two wavelengths at 852~nm and 780~nm. For this purpose, the visibility of the Rayleigh-scattered pattern for all three wavelengths is maximized iteratively. When trying to further optimize the individual visibilities for each wavelength by using additional Berek compensators in front of non-polarizing beamsplitters, the visibility could not be further increased. We therefore conclude that, after the optimization procedure, the single Berek compensator compensates all parasitic birefringence to zeroth order.

\begin{figure}
\centering
	\includegraphics[width=0.35\textwidth]{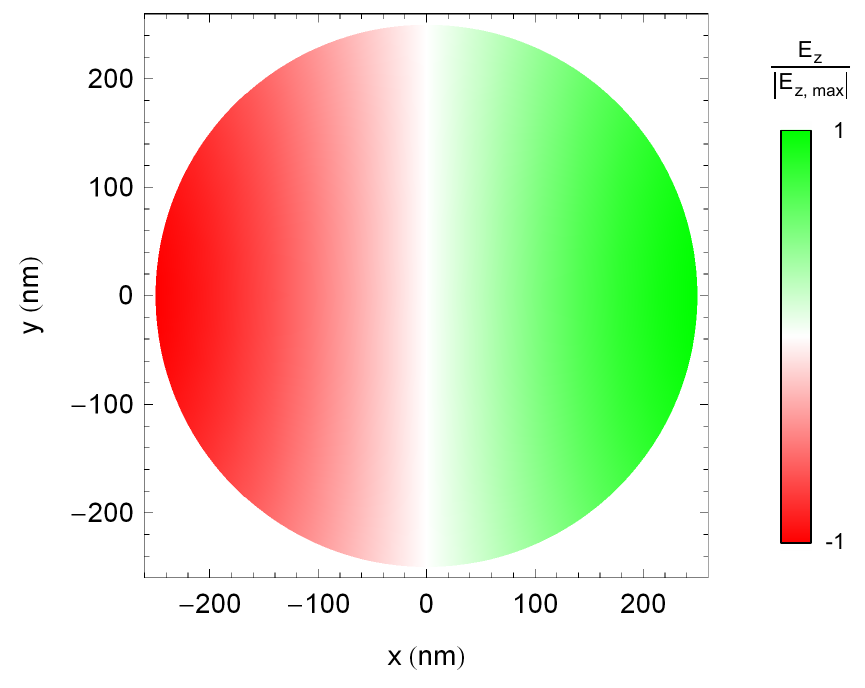}
	\caption{Plot of the $z$-component of the electric field of a HE$_{11}$ mode, quasi-linearly polarized along $x$, inside the 500-nm diameter nanofiber. The phase of oscillation is chosen to yield maximal field values. It is equal at all points with $x>0$ and opposed to the phase at the corresponding points with $x<0$. The vacuum wavelength of the guided light is assumed to be 852~nm.}
	\label{fig:PolVector}
\end{figure}

It is instructive to further consider the $\varphi$-dependence of the CCD signals when adjusting the polarizers to pass the $z$-component. Interestingly, in this case, we also observe a modulation with a contrast of 87~\%. This behavior can be understood when taking into account the spatial dependence of the $z$-component of a HE$_{11}$ mode, quasi-linearly polarized along $x$ (see Fig.~\ref{fig:PolVector}). As is apparent from this plot, the $z$-component of the E-field changes sign under reflection at the $y$-$z$-plane. As a consequence, when observing the scattered intensity along the $y$-direction, full destructive interference takes place. When observing along the $x$-direction, however, there is a non-zero path length difference to the detector between corresponding points which oscillate in phase opposition, thereby turning the destructive interference into partially constructive interference. Finally, when entirely removing the polarization filter of camera 1, the observed signal (plotted in blue) corresponds to the sum of the $y$- and $z$-signal. In this case, the terms $\propto \sin^2(\varphi - \varphi_0)$ and $\propto \cos^2(\varphi - \varphi_0)$ approximately sum up a constant offset on top of which the effect of the $\propto \sin^2(2(\varphi - \varphi_0))$ term is clearly apparent. This nicely illustrates the importance of filtering out the $z$-component when optimizing the polarization with our method.

\section{Temperature Measurements}
With the trapping laser polarizations optimized as outlined, the nanofiber-based trap is ready to be loaded. For this purpose, a magneto-optical trap (MOT) is superposed with the nanofiber. The MOT performs normally in spite of the presence of the 500-nm diameter fiber including the trapping fields. This is in agreement with the observations in \cite{Russell12}, where sub-Doppler temperatures have been measured for a MOT cloud of $^{85}$Rb atoms which was superposed with a bare nanofiber of 700~nm diameter. Both, the MOT loading as well as the transfer of the atoms to the nanofiber-based trap is performed like in Ref.~\cite{Vetsch10} and corresponds to a typical loading sequence of a steep optical dipole trap~\cite{Kuppens00}. The total number of atoms in the trap is determined from an absorption measurement using fiber-guided probe light resonant with the D2 line of Cesium. From the total power scattered by the trapped Cesium atoms at full saturation, their number is determined to be around 2000~\cite{Vetsch10}.

The knowledge of the temperature of the trapped atoms is relevant for numerous applications of our nanofiber trap. As an example, the thermal motion, combined with a position-dependent light shift on the atomic transition frequencies, leads to inhomogeneous broadening, thereby limiting the reversible dephasing time of quantum superpositions \cite{Kuhr05}. We will see below that in our case we load at most one atom per trapping site into the nanofiber trap due to the collisional blockade effect \cite{Schlosser02}. Under these conditions, a versatile method for measuring the temperature has been presented in Refs.~\cite{Alt03, Alt03E2}. It relies on the adiabatic lowering of the trapping potential and the measurement of the number of remaining trapped atoms as a function of the reduced trap depth. From an initial depth $U_0$, the trapping potential is lowered to a value $U_{\rm low}$. In this case, only the fraction of atoms with a certain maximum energy $E$ remains trapped. For adiabatic lowering, the action integral over one oscillation period remains conserved~\cite{LandauMech}. Therefore, in this case, the initial energy $E_0$ is related to the trap depth in the moment of the atom loss $U_{\rm esc}$ by a one-to-one correspondence. 

\begin{figure}
	\includegraphics{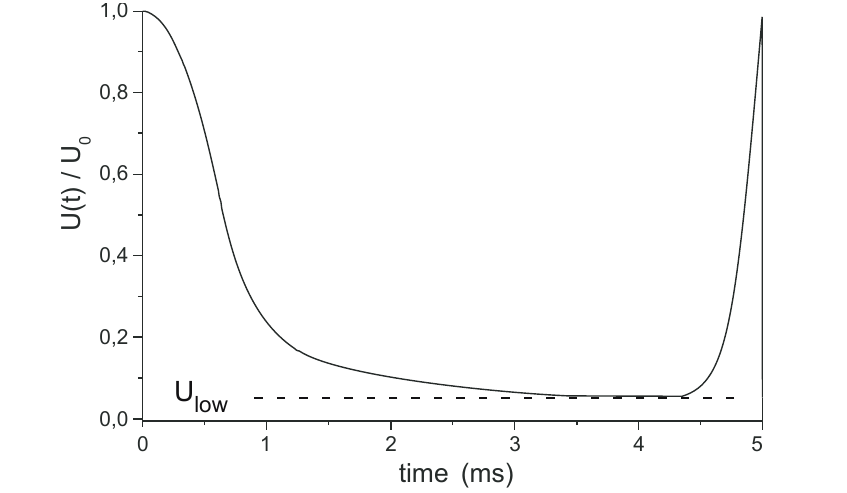}
	\caption{Plot of the adiabatic variation of the trap depth $U(t)$, normalized to the initial trap depth $U_0$, as a function of time~\cite{Alt03, Alt03E2}. $U_{\rm low}$ denotes the minimal trap depth reached during the whole adiabatic process.}
	\label{fig:Ramp}
\end{figure}
In order to experimentally fulfill the adiabaticity criterion, the optical dipole potential has to be changed on time scales that are longer than the largest oscillation period. An optimized experimental sequence to do so has been found in Ref.~\cite{Alt03} and is also applied here. The corresponding variation of the trap depth $U(t)$ is depicted in Fig.~\ref{fig:Ramp}. For the experimental implementation of the temperature measurement, the atoms are loaded from the MOT into the nanofiber trap with an initial depth of $U_0=\unit[400]{\mu K}$. Then, the trap is lowered by reducing the power of the red-detuned laser with an acousto-optical modulator (AOM) according to Fig.~\ref{fig:Ramp}. In order to determine the fraction of remaining atoms, the fiber-trapped ensemble is probed before and after the trap depth modification sequence, using a weak, detuned probe pulse of 500-$\mu$s duration.

\begin{figure}
	\includegraphics{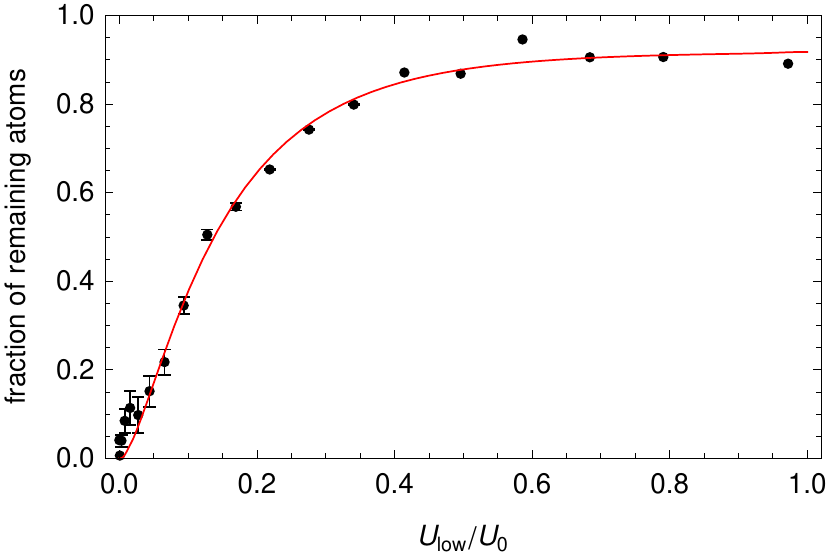}
	\caption{Fraction of atoms remaining in the trap as a function of the minimal trap depth $U_{\rm low}$. 
The data is fitted with an integrated Boltzmann distribution (see Eq.~(\ref{eq:intBoltz})), shown as a solid red line (see text for details). This fit yields a temperature of the atomic ensemble of $T=\unit[29.8\pm0.9]{\mu K}$.}
	\label{fig:survival}
\end{figure}

\begin{figure}
	\includegraphics{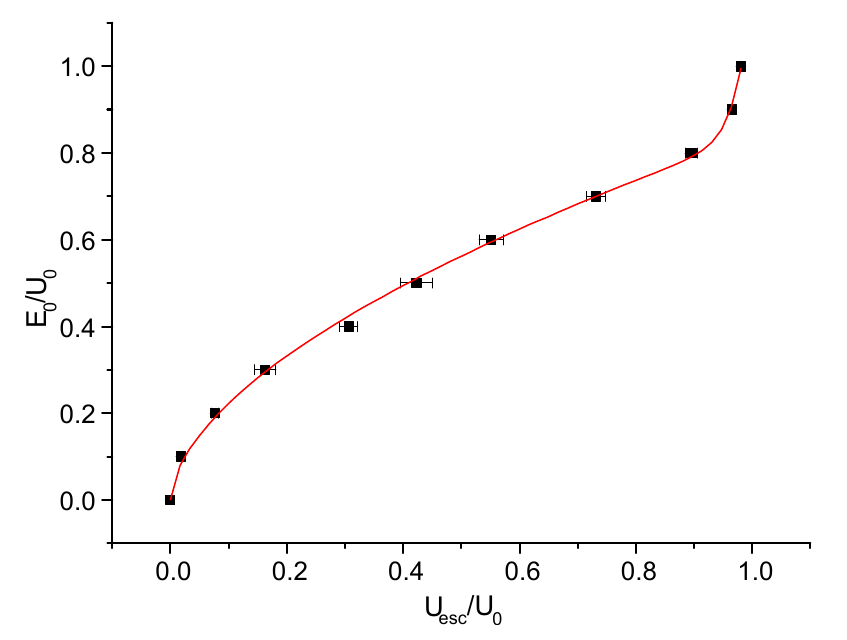}
	\caption{Initial energy $E_0$ of the atoms as a function of the escape trap depth $U_{\rm esc}$, both normalized to the initial trap depth $U_0$. The black data points are obtained from a Monte-Carlo simulation of $10^4$ classical trajectories of atoms trapped in one lattice site. The red curve is a polynomial fit.}
	\label{fig:E0Uesc}
\end{figure}

Figure~\ref{fig:survival} shows the corresponding experimental results. The maximum survival probability for a constant $U(t)=U_0$ reaches about 90~\%, in agreement with the trap lifetime of about 50~ms \cite{Vetsch10}. As expected, for decreasing $U_{\rm low}$, the fraction of remaining atoms in trap diminishes, reaching zero for vanishing trap depth. In order to derive the initial temperature of the atoms from the above measurement, the correspondence between $E_0$ and $U_{\rm esc}$ has to be determined. For this purpose, a Monte-Carlo simulation of classical atomic trajectories is performed since no analytical expression is known. Starting with different random initial positions in the trap and different random phases of the in-trap oscillation, the atomic dynamics is simulated. The trap parameters are chosen as in the experiment (see Table~\ref{tab:Trap} for the trap frequencies and Ref.~\cite{Vetsch10} for the detailed optical potentials). As a result of the simulation, the probability $p(E_0, U_{\rm low})$ for atoms of initial energy $E_0$ to remain trapped is obtained for different $U_{\rm low}$. For a given value of $E_0$ and in the case of infinitely slow lowering of the potential, $p(E_0, U_{\rm low})$ would correspond to a step function of $U_{\rm low}$. Any deviation from perfect adiabaticity will wash out the step-like behavior and $p(E_0, U_{\rm low})$ is well-described by an error function. In the simulation, the deviation from perfect adiabaticity occurs due to the finite size time steps which are used. From the resulting error function, the escape potential depth $U_{\rm esc}$ is thus defined by 
\begin{equation}
p(E_0, U_{\rm low}=U_{\rm esc})=0.5~.
\label{eq:Komma5}
\end{equation}
The obtained correspondence between $E_0$ and $U_{\rm esc}$ is shown in Fig.~\ref{fig:E0Uesc}. For the evaluation of the experimental results, the simulated data is fitted with a polynomial of the from $y=ax^b+cx^d$, shown as a solid red line.

In order to derive the fraction of remaining atoms for a given initial temperature $T$ as a function of $U_{\rm low}$, we model the initial cumulative energy distribution of the atomic ensemble in the nanofiber trap by an integrated three-dimensional Boltzmann distribution~\cite{Bagnato87},
\begin{equation}
P(E_0, T) = \int _0 ^{E_0} \frac{g(E)}{Z(T)} \exp \left( -\frac{E}{k_{\rm B} T} \right) \, dE~,
\label{eq:intBoltz}
\end{equation}
\begin{equation}
g(E) = \frac{2 \pi (2M)^{3/2}}{h^3} \; \int _{V(E)} \sqrt{E-U(\bf{r})} \, d^3r~,
\label{eq:DOS}
\end{equation}
where $g(E)$ denotes the density of states, $Z(T)$ is a normalization function, $M$ the atomic mass of Cesium, and $V(E)$ is the space available for particles with energy $E$. In the limit of $E\ll U_0$, the integral in Eq.~(\ref{eq:DOS}) can be approximately solved and Eq.~(\ref{eq:intBoltz}) can then be analytically integrated. This initial cumulative energy distribution $P(E_0, T)$ determines the fraction of remaining atoms after lowering the trap because atoms with an initial energy lower than $E_0(U_{\rm esc})$ will not leave the trap during the lowering sequence. Here, the mapping between $E_0$ and $U_{\rm esc}$ is made using the polynomial fit from Fig.~\ref{fig:E0Uesc}. Finally, in order to account for the finite trap lifetime, we multiply the resulting ideal survival probability $P(U_{\rm esc}, T)$ with a factor $p_{\rm max}$. The resulting fit function is shown as a solid red line in Fig.~\ref{fig:survival}, yielding $p_{\rm max}=92$~\%.

The temperature of the atomic ensemble in the nanofiber trap is found to be $k_{\rm B} T= 0.075(2) U_0$ which corresponds to $T=\unit[29.8 \pm 0.9]{\mu K}$ when using $U_0$ from Table~\ref{tab:Trap}. This value is clearly below the Doppler temperature and indicates the presence of sub-Doppler cooling mechanisms~\cite{Dalibard89, Metcalf99}. This is remarkable given the fact that it was far from evident that the MOT and the optical molasses would cool the atoms at all at a distance of only $\approx 200$~nm from the nanofiber surface. It leads us to conclude that the perturbation of the cooling laser beams is less detrimental to cooling than what might be expected from the modification of the fields by the nanofiber~\cite{LeKien09c}.

Using the obtained temperature value as well as the trap frequencies from Table~\ref{tab:Trap}, the mean occupation numbers of the vibrational levels in the trapping potential can be calculated. According to Ref.~\cite{Wineland79}, the occupation number is given by
\begin{equation}
\langle n_i \rangle = \frac{1}{\exp \left( h \nu_i / k_{\rm B} T \right) - 1},
\end{equation}
yielding $\langle n_z \rangle = 1.4$, $\langle n_r \rangle = 2.5$, and $\langle n_\varphi \rangle = 3.7$. The localization of an atom in the trap can be obtained from the full $1/\sqrt{e}$-width, $2\sigma_i$, of the probability distribution of the harmonic oscillator wave functions. For the parameters in this experiment, each atom is confined to a volume $V=2\sigma_z \times 2\sigma_r \times 2r\sigma_\varphi = (42 \times 67 \times 95) \unit[]{nm^3}=2.7\cdot 10^{-16}$~cm$^3$ at its lattice site. When probing the atoms with resonant light propagating through the nanofiber, the interaction thus takes place in the Lamb-Dicke regime (Lamb-Dicke parameter: $\eta=0.08$), thereby fulfilling the necessary condition for resolved sideband cooling to work efficiently.

\section{Atom-number distribution}
The large trap frequencies realized in the present set-up (see Table~\ref{tab:Trap}), combined with the scattering of near resonant light during transfer of the atoms from the MOT to the nanofiber trap, realize a situation in which the collisional blockade effect might occur and lead to a sub-Poissonian atom number distribution in each trapping site \cite{Schlosser02}. In particular, over a large range of parameters, the maximum atom number per trapping site can be limited to one while the average occupation is close to 0.5. The necessary condition for observing such a collisional blockade is to have the one-body loss rate during the loading process, $\gamma$, much smaller than the light-induced two-body loss rate, $\beta/V$, where $V$ is the trapping volume \cite{Schlosser02}. We experimentally observe $\gamma\approx 1$~s$^{-1}$, consistent with loss due to background gas collisions at the base pressure in our set-up. Using the volume inferred from the temperature measurement (see above) and a typical value of $\beta=10^{-10}$~cm$^3$~s$^{-1}$ for Cesium atoms~\cite{Ueberholz00}, we find $\beta/V=3.7\cdot 10^5$~s$^{-1}\gg \gamma$. If, in addition, the loading rate of the nanofiber trap, $R$, fulfills the condition
\begin{equation}
\gamma/2 < R < \beta/4V~,
\label{eq:collblock}
\end{equation}
collisional blockade should thus occur \cite{Schlosser02}. It is, however, not trivial to determine the loading rate $R$ from a priori considerations or independent measurements. For this reason we give an estimation of $R$ in the following. The loading rate in our set-up clearly exceeds the lower boundary in condition (\ref{eq:collblock}) because we load about 2000 atoms into approximately 4000 trapping sites in a loading sequence of $\unit[50]{ms}$ duration. In order to estimate an upper boundary of $R$, we compare our trapping sites with the tightly focused single-spot dipole trap in Ref.~\cite{Schlosser02} where the maximum loading rate was smaller than 1000~s$^{-1}$ when loading from a standard MOT. In our case, both the trapping volume and the solid acceptance angle for loading the trap are significantly smaller than in Ref.~\cite{Schlosser02}. From these simple considerations, we estimate to realize a loading rate much smaller than the upper boundary in condition (\ref{eq:collblock}). Consequently, we expect our trap to operate in the collisional blockade regime.

In order to experimentally check the occurrence of the collisional blockade, we load the nanofiber trap as described in \cite{Vetsch10}, excite the atoms for 2~ms with near resonant fiber-guided probe light, and image their fluorescence using an EMCCD camera. The latter is oriented as indicated in Fig.~\ref{fig:setup}(a), i.e., with the observation axis perpendicular to the nanofiber axis. In order to limit heating due to photon recoil, the probe light power is set to $\unit[500]{pW}$ and its frequency is red-detuned by $\unit[20]{MHz}$ with respect to the AC-stark-shifted $F=4 \rightarrow F^\prime=5$ transition of the Cesium D2 line. The plane of the quasi-linear polarization of the probe light is parallel to the plane containing the atoms and orthogonal to the observation axis. In order to obtain a good signal to noise ratio, this measurement is repeated in 320 consecutive experimental runs and the background-corrected EMCCD fluorescence  images are averaged.

\begin{figure}
	\includegraphics{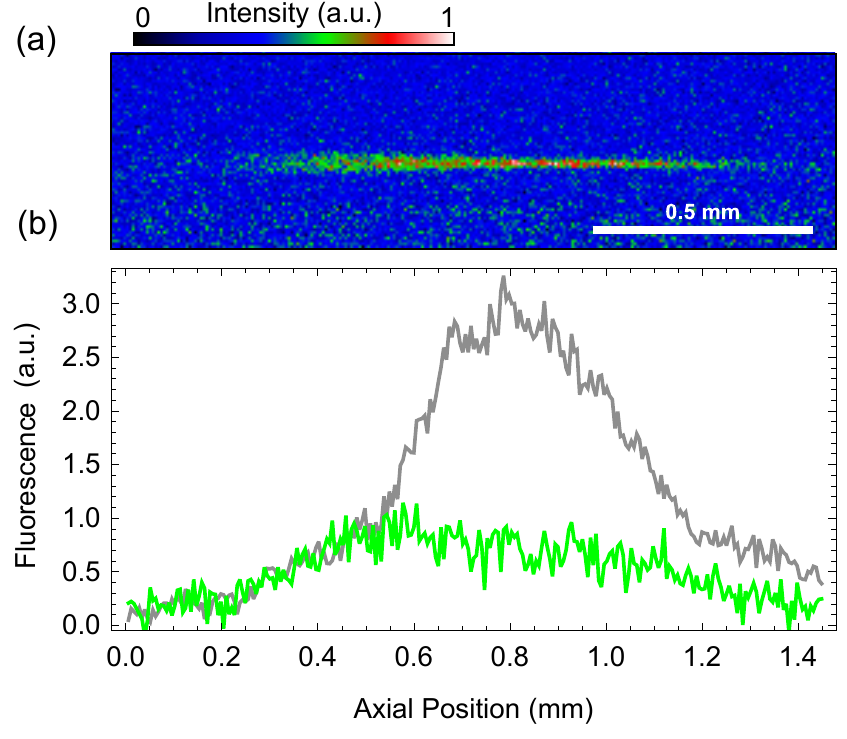}
	\caption{(a) Background corrected fluorescence image of the nanofiber-trapped atoms, recorded with an EMCCD camera and averaged over 320 exposures. (b, solid green line) The same signal, integrated perpendicular to the nanofiber axis. In addition, the fluorescence of the MOT prior to loading the nanofiber trap is shown as a solid gray line. In order to allow a comparison of the fluorescence distributions, the MOT-signal has been rescaled.}
	\label{fig:Fluo}
\end{figure}
The resulting image is shown in Fig.~\ref{fig:Fluo}~(a), while Fig.~\ref{fig:Fluo}~(b) shows the signal integrated perpendicular to the nanofiber axis as a solid green line. The latter fluorescence signal of the nanofiber-trapped atoms is higher in the left part of the graph than in the right part. This observation can be easily understood: the probe laser enters the fiber from the left in Fig.~\ref{fig:Fluo} and gets partially absorbed starting with the leftmost trapped atoms. Consequently, the probe laser intensity $I_P(z)$ varies along the fiber. Denoting the line number density of the atomic ensemble as $\rho (z)$, the fluorescence intensity $I_F(z)$ will thus be given by
\begin{equation}
I_F(z) \propto \rho (z) I_P(z) =\rho (z) I_0 \exp \left( -\frac{\sigma}{A_{\rm eff}} \int _0 ^z \rho(z') dz' \right)~,
\label{eq:lambert}
\end{equation}
where $\sigma$ is the atomic absorption cross section and $A_{\rm eff}$ is the effective cross sectional area of the nanofiber guided mode at the probe wavelength~\cite{Warken07}. The settings of the probe light intensity and detuning correspond to a saturation parameter $s < 1/10$ at the position of the atoms. This justifies the application of Lambert-Beer's law in Eq.~(\ref{eq:lambert}).

Figure~\ref{fig:Fluo}~(b) also shows the fluorescence signal from the MOT cloud prior to loading the trap (solid gray line). It is well approximated by a Gaussian with a full $1/\sqrt{e}$-width of $2\sigma_{\rm MOT}=0.42$~mm. In the following, we assume that the MOT density distribution and thus also the loading rate of the nanofiber trap, $R$, follow this Gaussian shape. In order to model the nanofiber fluorescence signal in Fig.~\ref{fig:Fluo}~(b) using Eq.~(\ref{eq:lambert}), we now have to derive the line number distribution of the fiber trapped atoms $\rho (z)$ from the loading rate
\begin{equation}
R(z)=R_{\rm max} \exp \left[ -(z-z_0)^2/2\sigma_{\rm MOT}^2 \right]~,
\end{equation}
where $z_0$ denotes the center of the MOT.

\begin{figure}
	\includegraphics{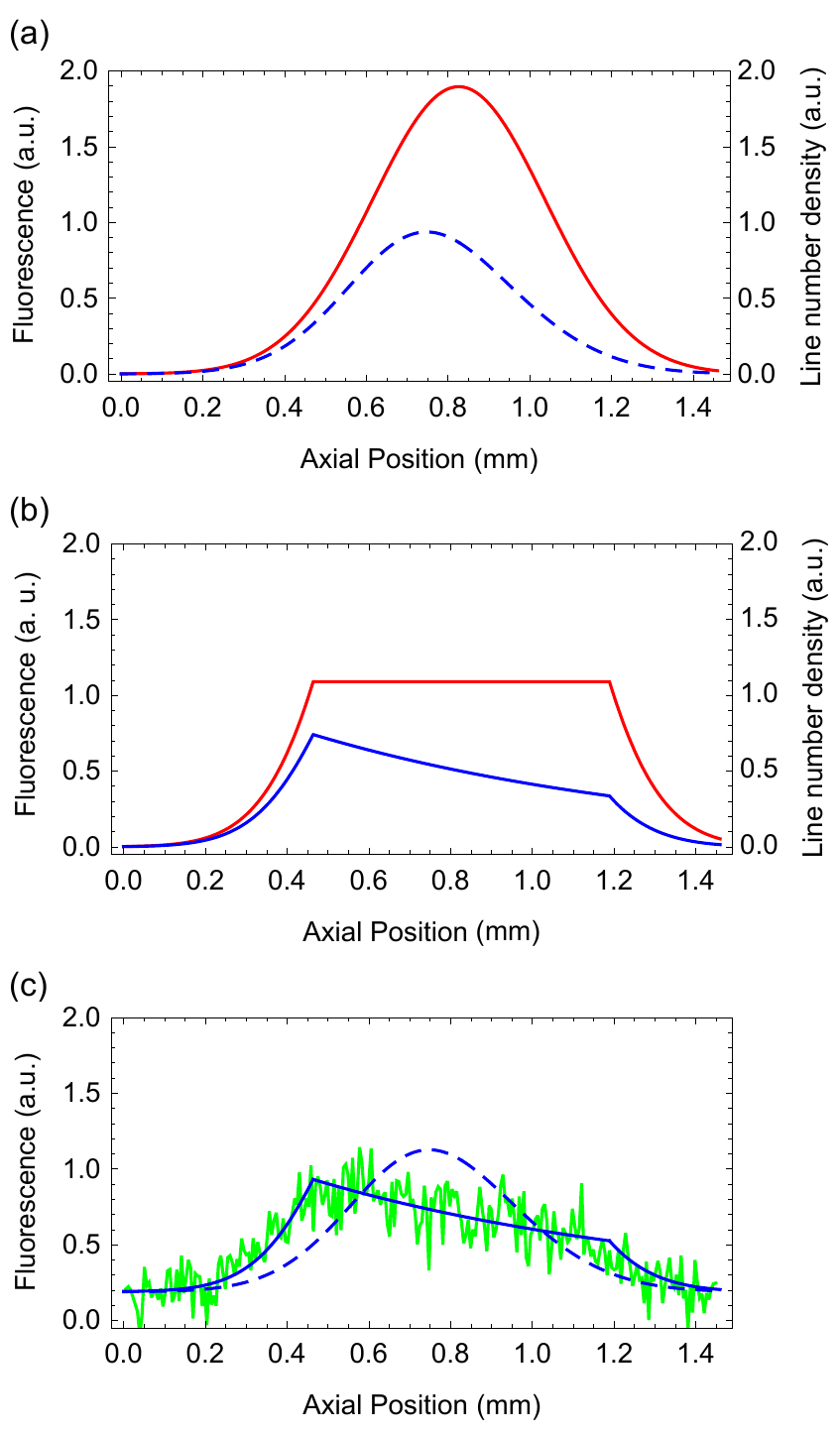}
	\caption{(a) Modeled line number density (solid red line) and resulting fluorescence signal (dashed blue line) of the nanofiber-trapped atoms in the absence of the collisional blockade effect. (b) Modeled line number density (solid red line) assuming collisional blockade and corresponding fluorescence signal (solid blue line). The threshold value for clipping the line number density is chosen to occur at 23~\% of the maximum loading rate, $R_{\rm \max}$, in order to yield the best agreement with the experimentally observed fluorescence signal. (c) Comparison of the experimentally observed fluorescence signal (solid green line) with the modeled fluorescence signals from (a) and (b) including an offset to account for a residual background. The experimental data shows very good agreement with the prediction derived under the assumption of collisional blockade.}
	\label{fig:theofluo}
\end{figure}
Given that $R_{\rm max}$ is unknown, it is unclear whether $R(z)$ enters the collisional blockade regime. For this reason, we first assume that this is not the case and that $\rho(z)$ is simply proportional to $R(z)$. Furthermore, from an independent measurement, we know that the overall optical depth of the nanofiber-trapped atomic ensemble for the probe settings is about ${\rm OD}=1$, thereby providing a normalization condition for the integrated line number density $\int \rho(z^\prime)\, dz^\prime$. Figure~\ref{fig:theofluo}~(a) shows $\rho(z)$ under these assumptions (solid red line) as well as the resulting fluorescence signal (dashed blue line) predicted according to Eq.~(\ref{eq:lambert}). As can be seen in Fig.~\ref{fig:theofluo}~(c), this predicted signal does not reproduce the experimental data.

We therefore assume that $\rho(z)$ corresponds to a truncated Gaussian as would be expected if collisional blockade occurred. For simplicity, we model this truncation by a hard clipping of the Gaussian as shown by the solid red line in Fig.~\ref{fig:theofluo}~(b). In this case, the predicted fluorescence signal is in very good qualitative and quantitative agreement with the experiment, see solid blue line in Fig.~\ref{fig:theofluo}~(c). We interpret this observation as an experimental proof of a collisional blockade effect in our experiment. This claim is supported by the following estimation: For the collisional blockade, an average filling factor of 0.5 atoms per trapping site is expected. In our trapping configuration, we have four trapping sites per micrometer along the nanofiber. For the sample length of $\sim 1$~mm, we would thus expect $\sim 2000$ trapped atoms. This number is in good agreement with the atom number inferred from the saturation measurement in Ref.~\cite{Vetsch10}.

\section{Conclusions}
We presented a practical {\em in situ} polarization analysis of light fields guided in an optical nanofiber and showed that this method can be employed for the optimization of the polarization configuration of our nanofiber-based atom trap. Moreover, we carried out an in-depth study of the trap loading performance using a magneto-optical trap as an atom source. We found that our nanofiber trap yields comparable results, i.e., sub-Doppler temperatures and collisional blockade, as in steep free-beam optical dipole traps and optical lattices. This shows that the possible perturbations of the laser-cooling and loading process, introduced by the presence of the nanofiber, are not significant. In this respect, the nanofiber trap is thus as simple to operate as a standard trap. At the same time, however, it realizes an efficient optical interface for the trapped atoms. Combined with the single-atom occupancy of the trapping sites, this therefore opens a realm of applications for nanofiber-trapped atomic ensembles.

\section*{Acknowledgments}
This work was supported by the Volkswagen Foundation (Lichtenberg Professorship), the European Science Foundation (EURYI Award), and the Austrian Science Fund (CoQuS Graduate school, project W1210-N16). The authors wish to thank LOT-Oriel for the loan of the EMCCD camera.

%%%%%%%%%%%%%%%%%%%%%%%%%%%%%%%%%%%%%%%%%%%%%%%%%%%%%%%%%%%%%%%%%%%%%%%%%%%%%%%%%%%%%%%
%%%%%%%%%%%%%%%%%%%%%%%%%%%%%%%%%%%%%%%%%%%%%%%%%%%%%%%%%%%%%%%%%%%%%%%%%%%%%%%%%%%%%%%
%%%%%%%%%%%%%%%%%%%%%%%%%%%%%%%%%%%%%%%%%%%%%%%%%%%%%%%%%%%%%%%%%%%%%%%%%%%%%%%%%%%%%%%

\bibliographystyle{IEEEtran}
%\bibliographystyle{mzsty}


%\bibliography{FF}% Produces the bibliography via BibTeX.

\begin{thebibliography}{}
\providecommand{\url}[1]{#1}
\csname url@samestyle\endcsname
\providecommand{\newblock}{\relax}
\providecommand{\bibinfo}[2]{#2}
\providecommand{\BIBentrySTDinterwordspacing}{\spaceskip=0pt\relax}
\providecommand{\BIBentryALTinterwordstretchfactor}{4}
\providecommand{\BIBentryALTinterwordspacing}{\spaceskip=\fontdimen2\font plus
\BIBentryALTinterwordstretchfactor\fontdimen3\font minus
  \fontdimen4\font\relax}
\providecommand{\BIBforeignlanguage}[2]{{%
\expandafter\ifx\csname l@#1\endcsname\relax
\typeout{** WARNING: IEEEtran.bst: No hyphenation pattern has been}%
\typeout{** loaded for the language `#1'. Using the pattern for}%
\typeout{** the default language instead.}%
\else
\language=\csname l@#1\endcsname
\fi
#2}}
\providecommand{\BIBdecl}{\relax}
\BIBdecl

\end{thebibliography}


\begin{thebibliography}{10}
\providecommand{\url}[1]{#1}
\csname url@samestyle\endcsname
\providecommand{\newblock}{\relax}
\providecommand{\bibinfo}[2]{#2}
\providecommand{\BIBentrySTDinterwordspacing}{\spaceskip=0pt\relax}
\providecommand{\BIBentryALTinterwordstretchfactor}{4}
\providecommand{\BIBentryALTinterwordspacing}{\spaceskip=\fontdimen2\font plus
\BIBentryALTinterwordstretchfactor\fontdimen3\font minus
  \fontdimen4\font\relax}
\providecommand{\BIBforeignlanguage}[2]{{%
\expandafter\ifx\csname l@#1\endcsname\relax
\typeout{** WARNING: IEEEtran.bst: No hyphenation pattern has been}%
\typeout{** loaded for the language `#1'. Using the pattern for}%
\typeout{** the default language instead.}%
\else
\language=\csname l@#1\endcsname
\fi
#2}}
\providecommand{\BIBdecl}{\relax}
\BIBdecl

\bibitem{Vetsch10}
E.~Vetsch, D.~Reitz, G.~Sagu\'e, R.~Schmidt, S.~T. Dawkins, and
  A.~Rauschenbeutel, ``Optical interface created by laser-cooled atoms trapped
  in the evanescent field surrounding an optical nanofiber,'' \emph{Phys. Rev.
  Lett.}, vol. 104, no.~20, p. 203603, 2010.

\bibitem{Dawkins11}
S.~T. Dawkins, R.~Mitsch, D.~Reitz, E.~Vetsch, and A.~Rauschenbeutel,
  ``Dispersive optical interface based on nanofiber-trapped atoms,''
  \emph{Phys. Rev. Lett.}, vol. 107, p. 243601, 2011.

\bibitem{LeKien04}
F.~Le~Kien, V.~I. Balykin, and K.~Hakuta, ``Atom trap and waveguide using a
  two-color evanescent light field around a subwavelength-diameter optical
  fiber,'' \emph{Phys. Rev. A}, vol.~70, no.~6, p. 063403, 2004.

\bibitem{Rauschenbeutel11}
A.~Rauschenbeutel, S.~T. Dawkins, R.~Mitsch, D.~Reitz, G.~Sagu\'e, R.~Schmidt,
  and E.~Vetsch, ``Trapping and interfacing cold neutral atoms using optical
  nanofibers,'' in \emph{Proceedings of the 20th ICOLS}, W.~Ertmer and
  R.~Scholz, Eds., 2011, p. 209.

\bibitem{Hafezi11}
M.~Hafezi, D.~E. Chang, V.~Gritsev, E.~A. Demler, and M.~D. Lukin, ``Photonic
  quantum transport in a nonlinear optical fiber,'' \emph{Europhys. Lett.},
  vol.~94, no.~5, p. 54006, 2011.

\bibitem{Kuppens00}
S.~J.~M. Kuppens, K.~L. Corwin, K.~W. Miller, T.~E. Chupp, and C.~E. Wieman,
  ``Loading an optical dipole trap,'' \emph{Phys. Rev. A}, vol.~62, no.~1, p.
  013406, 2000.

\bibitem{Schlosser02}
N.~Schlosser, G.~Reymond, and P.~Grangier, ``Collisional blockade in
  microscopic optical dipole traps,'' \emph{Phys. Rev. Lett.}, vol.~89, no.~2,
  p. 023005, 2002.

\bibitem{LeKien04b}
F.~Le~Kien, J.~Liang, K.~Hakuta, and V.~Balykin, ``Field intensity
  distributions and polarization orientations in a vacuum-clad
  subwavelength-diameter optical fiber,'' \emph{Opt. Comm.}, vol. 242, no. 4-6,
  p. 445, 2004.

\bibitem{Yariv07}
A.~Yariv and P.~Yeh, \emph{Photonics - Optical Electronics in Modern
  Communication}.\hskip 1em plus 0.5em minus 0.4em\relax Oxford University
  Press, 2007.

\bibitem{Russell12}
L.~Russell, K.~Deasy, M.~Daly, M.~Morrissey, and S.~N. Chormaic, ``Sub-doppler
  temperature measurements of laser-cooled atoms using optical nanofibres,''
  \emph{Meas. Sci. Technol.}, vol.~23, p. 015201, 2012.

\bibitem{Kuhr05}
S.~Kuhr, W.~Alt, D.~Schrader, I.~Dotsenko, Y.~Miroshnychenko,
  A.~Rauschenbeutel, and D.~Meschede, ``Analysis of dephasing mechanisms in a
  standing-wave dipole trap,'' \emph{Phys. Rev. A}, vol.~72, p. 023406, 2005.

\bibitem{Alt03}
W.~Alt, D.~Schrader, S.~Kuhr, M.~M\"uller, V.~Gomer, and D.~Meschede, ``Single
  atoms in a standing-wave dipole trap,'' \emph{Phys. Rev. A}, vol.~67, p.
  033403, 2003.

\bibitem{Alt03E2}
------, ``Erratum: Single atoms in a standing-wave dipole trap [phys. rev. a
  67, 033403 (2003)],'' \emph{Phys. Rev. A}, vol.~71, p. 019905, 2005.

\bibitem{LandauMech}
L.~D. Landau and E.~M. Lifshitz, \emph{Mechanics}, 3rd~ed.\hskip 1em plus 0.5em
  minus 0.4em\relax Butterworth-Heinemann, 1981.

\bibitem{Bagnato87}
V.~Bagnato, D.~E. Pritchard, and D.~Kleppner, ``Bose-einstein condensation in
  an external potential,'' \emph{Phys. Rev. A}, vol.~35, p. 4354, 1987.

\bibitem{Dalibard89}
J.~{Dalibard} and C.~{Cohen-Tannoudji}, ``{Laser cooling below the Doppler
  limit by polarization gradients: Simple theoretical models},'' \emph{J. Opt.
  Soc. Am. B}, vol.~6, p. 2023, 1989.

\bibitem{Metcalf99}
H.~J. Metcalf and P.~van~der Straten, \emph{Laser cooling and trapping}.\hskip
  1em plus 0.5em minus 0.4em\relax Springer Press, New York, 1999.

\bibitem{LeKien09c}
F.~Le~Kien and K.~Hakuta, ``Microtraps for atoms outside a fiber illuminated
  perpendicular to its axis: Numerical results,'' \emph{Phys. Rev. A}, vol.~80,
  no.~1, p. 013415, 2009.

\bibitem{Wineland79}
D.~J. Wineland and W.~M. Itano, ``Laser cooling of atoms,'' \emph{Phys. Rev.
  A}, vol.~20, p. 1521, 1979.

\bibitem{Ueberholz00}
B.~{Ueberholz}, S.~{Kuhr}, D.~{Frese}, D.~{Meschede}, and V.~{Gomer},
  ``{Counting cold collisions},'' \emph{J. Phys. B}, vol.~33, p. L135, 2000.

\bibitem{Warken07}
F.~Warken, E.~Vetsch, D.~Meschede, M.~Sokolowski, and A.~Rauschenbeutel,
  ``Ultra-sensitive surface absorption spectroscopy using sub-wavelength
  diameter optical fibers,'' \emph{Opt. Express}, vol.~15, no.~19, p. 11952,
  2007.

\end{thebibliography}

%%%%%%%%%%%%%%%%%%%%%%%%%%%%%%%%%%%%%%%%%%%%%%%%%%%%%%%%%%%%%%%%%%%%%%%%%%%%%%%%%%%%%%%%
%%%%%%%%%%%%%%%%%%%%%%%%%%%%%%%%%%%%%%%%%%%%%%%%%%%%%%%%%%%%%%%%%%%%%%%%%%%%%%%%%%%%%%%%
%%%%%%%%%%%%%%%%%%%%%%%%%%%%%%%%%%%%%%%%%%%%%%%%%%%%%%%%%%%%%%%%%%%%%%%%%%%%%%%%%%%%%%%%

\begin{biography}[{\includegraphics[width=1in,height=1.25in,clip,keepaspectratio]{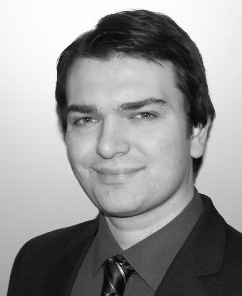}}]{Eugen Vetsch}
was born in Dushanbe, Tajikistan, in 1978. He graduated from the University of Bonn, Germany, in 2006 and received the Ph.D.~degree in physics, in 2011, from the University of Mainz, Germany. His main research activities have been nanofiber-based optical trapping and interfacing of cold atoms.

He took part in development of industrial attenuation measurements techniques for UV-silica fibers at Heraeus Quartzglas GmbH. Currently, he is occupationally involved in the research and development of Head-Up Displays at Continental Automotive GmbH.
He is a member of the German Physical Society (DPG).
\end{biography}

\begin{biography}[{\includegraphics[width=1in,height=1.25in,clip,keepaspectratio]{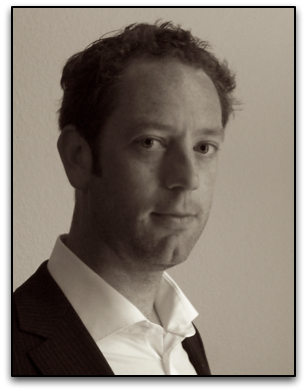}}]{Samuel T.~Dawkins}
was born in Perth, Western Australia, in 1976. He received a Ph.D. degree in physics from the University of Western Australia in 2007. Thereafter, he helped develop a mercury-based optical lattice clock at Syst\`{e}mes de R\'{e}f\'{e}rence Temps Espace (SYRTE) at the Paris Observatory. Since 2009, he has worked in the field of quantum optics at the Johannes Gutenberg-University in Mainz, Germany, where he has helped develop optical nanofiber-based atom traps and segmented ion-traps, both prospective tools for the manipulation of quantum information.
\end{biography}

\begin{biography}[{\includegraphics[width=1in,height=1.25in,clip,keepaspectratio]{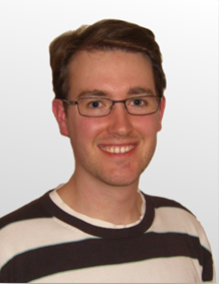}}]{Rudolf Mitsch}
was born in Heppenheim, Germany, in 1983. He received the B.Sc. degree in physics in 2006, and the M.Sc.  degree in physics in 2009 from the Technical University Darmstadt in Germany, both in the group of Prof. Thomas Walther, where he worked on a laser system for trapping cold mercury atoms. In 2009 he started to work toward the Ph.D.~degree in the group of Prof. Arno Rauschenbeutel at the Johannes Gutenberg-University Mainz, Germany. In 2010 he moved with the group to Vienna, Austria. Here, he is affiliated with the Atominstitut at the Vienna University of Technology and the Vienna Center for Quantum Science and Technology. Rudolf Mitsch currently acknowledges a CoQuS Ph.D.~scholarship for the work on fiber-based atom traps.
\end{biography}

\begin{biography}[{\includegraphics[width=1in,height=1.25in,clip,keepaspectratio]{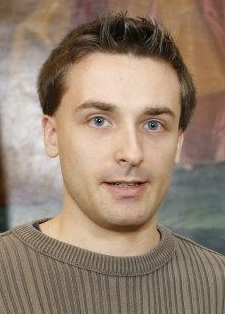}}]{Daniel Reitz}
was born in Hadamar, Germany, in 1982. He studied physics at the Johannes Gutenberg-University Mainz, Germany, and received his diploma degree in 2008 in the group of Prof. Herwig Ott, where he worked on resolving single sites of an optical lattice with an electron microscope. After that, he joined the group of Prof. Arno Rauschenbeutel at the Johannes Gutenberg-University in Mainz and started to work toward the Ph.D.~degree in the field of nanofiber quantum optics. In 2010 he moved with the group to Vienna, Austria, where he is affiliated with the Atominstitut at the Vienna University of Technology and the Vienna Center for Quantum Science and Technology. In Vienna, Daniel Reitz continues his work on the fiber-based atom trap.
\end{biography}

\begin{biography}[{\includegraphics[width=1in,height=1.25in,clip,keepaspectratio]{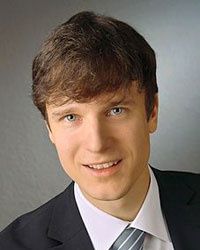}}]{Philipp Schneeweiss} was born in Potsdam, Germany, in 1981. He studied physics at the Technical University of Dresden (graduation as MSc.~in 2006), Germany, and at the University of Calgary, Canada. In 2011, he obtained a Ph.D.~degree from the University of T\"ubingen, Germany, with a work on ultracold atoms interacting with carbon nanotubes. After that, he joined, as a Postdoctoral Researcher, the group of A.~Rauschenbeutel at the Atominstitut at the Vienna University of Technology, Austria. Dr.~Schneeweiss is also affiliated with the Vienna Center for Quantum Science and Technology. His main area of research is cold atoms interfaced with optical nanofibers.
\end{biography}

\begin{biography}[{\includegraphics[width=1in,height=1.25in,clip,keepaspectratio]{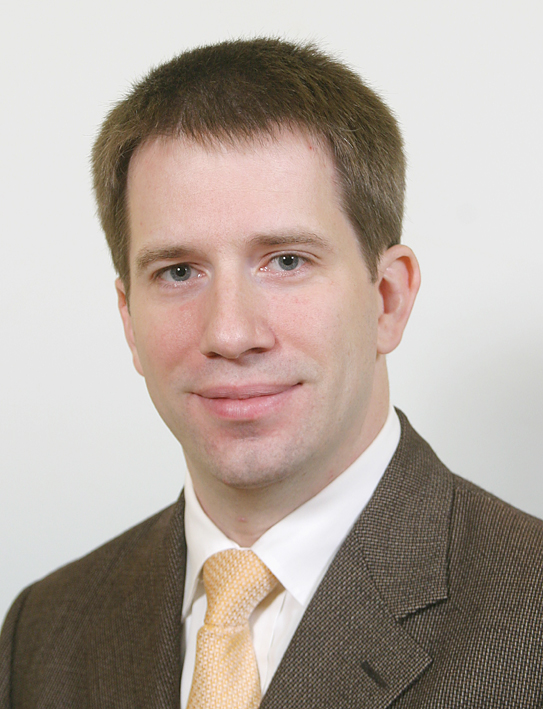}}]{Arno Rauschenbeutel} was born in D\"usseldorf, Germany, in 1971. He did his Ph.D. in experimental quantum optics at the Laboratoire Kastler Brossel of the Ecole normale sup\'{e}rieure, Paris, France, and obtained his Ph.D.~degree from the Universit\'{e} Paris~VI in 2001.

He worked as a Senior Scientist at the Institute for Applied Physics at the University of Bonn, as a Professor at the University of Mainz, and currently holds the Chair for Applied Quantum Physics at the Atominstitut at the Vienna University of Technology, Austria. Furthermore, he is one of the founding members of the Vienna Center for Quantum Science and Technology. His fields of interest include experimental quantum optics, nanophotonics, cavity quantum electrodynamics, optical nanofibers, optical microresonators, and cooling and trapping of neutral atoms.

Prof. Rauschenbeutel received a Marie Curie Excellence Award of the European Commission, a European Young Investigators Award of the European Science Foundation, and a Lichtenberg-Professorship of the Volkswagen-Foundation.

\end{biography}

%\includegraphics[width=1in,height=1.25in,clip,keepaspectratio]{Pics/test}
%\end{IEEEbiography}
%% insert where needed to balance the two columns on the last page with
%%\newpage


%\vfill


\end{document}